\documentclass[prd,aps,showpacs,superscriptaddress,nofootinbib,notitlepage,floatfix]{revtex4-2}

\usepackage{epsfig,graphicx,amsfonts,amsmath,amssymb}

\RequirePackage{mathrsfs}
\usepackage{color}
\usepackage{epsf}
\usepackage{epstopdf}

\def\eq#1{{Eq.~(\ref{#1})}}
\newcommand{\Le}{\left(}
\newcommand{\Ra}{\right)}

\newcommand{\beq}{\begin{equation}}
	\newcommand{\eeq}{\end{equation}}
\newcommand{\beqar}{\begin{eqnarray}}
	\newcommand{\eeqar}{\end{eqnarray}}
\newcommand{\D}{\partial}
\newcommand{\e}{{\rm e}}
\newcommand{\E}{{\rm E}}
\newcommand{\g}{{\rm g}}

\newcommand{\ph}{\varphi}

\newcommand{\om}{\omega}

\begin{document}
	\title{Dynamical signature: complex manifolds, gauge fields and non-flat tangent space }
	\author{S. Bondarenko}
	\affiliation{Ariel University, Ariel 4070000, Israel}

\date{\today}
	
\begin{abstract}
	
	 Theoretical possibilities of models of gravity with dynamical signature are discussed. The different scenarios of the signature change are proposed in the framework of 
Einstein-Cartan gravity. We consider, subsequently, the dynamical signature in the 	model of the complex manifold with complex coordinates and complex metric introduced, 
a complexification of the manifold and coordinates through new gauge fields, an additional gauge symmetry for the Einsten-Cartan vierbein fields and non-flat tangent space for  
the metric in the Einstein-Cartan gravity. 
A new small parameter, which characterizes a degree of the deviation of the signature from the background one, is introduced in all models. The zero value of this parameter corresponds to the
signature of an initial background metric.   
In turn, in the models with gauge fields present, this parameter represents a coupling constant of the gauge symmetry group. 
The mechanism of metric's determination through  induced gauge fields with defined signature in the corresponding models 
is considered. The ways of the signature change through the gauge fields dynamics are reviewed,
the consequences and applications of the proposed ideas are discussed as well.
\end{abstract}
	
	\maketitle

\section{Introduction}

 The idea of a metric with changing signature, looking unusual, attracts a lot of attention in the quantum cosmology and quantum gravity, see different 
aspects of the problem in \cite{MisWh,Hawk1,Sakh,Ander,Gero,Sork,Strom,Dray,Viss,Barv,borde,Kri,SigC,Green,Spont}. Whereas all experiments and observations do not question the fact that the
classical  metric of the  Universe has Lorentzian signature, 
these are two windows which we can not look through to check the signature. We do not know a lot about the quantum gravity world, there are 
theoretical models that allows change of the signature at the quantum level, see for example \cite{MisWh}. The very beginning of the Universe is an another corner which can hide the possible change of the signature, see for example \cite{Hawk1}.

  Among other, there are two parameters of the classical gravity which is very interesting to explore in particular. They are the number of space-time dimensions and signature of the metric, the last one is related as well to the arrow of time. 
The main purpose of the proposed article is an investigation of the signature issue. We consider mathematical possibilities of the formulation of the gravity formalisms with a metric which can change dynamically i.e.
we discuss possible scenarios of gravity with signature which takes on values in the field of complex numbers. 
Particularly, a determination of the signature as Euclidean or Lorentzian is happening in the models due to some additional mechanisms.
There are a few possibilities which will be discussed. The first one is a comlexification of the metric and manifold's coordinates. The gravity theories with complex metric introduced are not new of course, there is 
a complex metric calculation and use in \cite{Moffat} for example. The complex manifold also arises naturally in the twistor space formulation, see  \cite{Penrose,Plebanski} for the different examples 
of projection of complex Minkowski space  in order to describe gravitons outside of a linearized gravity framework. The examples of the introduction of the complex manifold in the strings frameworks sector
can be found in \cite{Witten, Str}  as well.  Nevertheless, in the present approach we consider the problem differently. There is no classical eight dimensional complex world around, therefore we introduce a small parameter which zero value corresponds to the real manifold and real coordinates. In turn, the non-zero value of the parameter adds  additional dimensions to the manifold as well as an additional complex part to the  metric tensor. These additional contributions to the four dimensional gravity are proportional to the powers of the parameter, therefore
the smallness of the parameter allows to establish a perturbative scheme related to the expansions of objects of interest with respect to it. We consider a factorization of the real four dimensional gravity from the additional four dimensional space of complex phases. If we assume that the small parameter of the problem is not small only at some special conditions then we get that the additional contributions are important only in these extremal cases, this issue we discuss in the next section. We do not consider a reduction of the complex manifold to the four dimensional real one, but instead 
we assume an coexisting of an additional contribution to the usual metric. The signature of this contribution is dynamical and it's value is limited by the value of the 
new parameter introduced.

  The natural next step in this direction is a complexification of the manifold with the help of new gauge fields. In this case the phases of the coordinates are defined by the gauge fields. The introduced parameter in this picture is a coupling constant of the corresponding gauge group. Considering the Einstein-Cartan gravity as the base of the approach, we, consequently, obtain an additional correction to the vierbein field which  depends on the introduced phases and is proportional to the new parameter. Therefore, the metric components acquire a phase factor which makes it's signature complex and non-determined in general.
Whereas the phases of the metric's components are defined by the gauge fields values, we discuss a possibility of a determination of any requested value of the signature with the help of 
induced values of the gauge fields which satisfy some boundary conditions. In this scenario, the signature's value is fixed by these induced fields. The mechanism can be defined in the case of 
a consideration of the overall metric signature as well as for the case of description of
a signature's  fluctuations above some background metric.

 The change of the signature in the approach through the gauge fields is considered in two different formulations in turn. As mentioned above, the first possibility is a complexification of the manifold and metric through a complexification of the coordinates achieved by the gauge fields. The other possibilities are related to the redefinition of the structure of the Einstein-Cartan gravity with the use of the new gauge fields, i.e. we consider possible generalizations of vierbein fields. The first possibility we consider is the simplest one, we complexify the vierbein by the gauge field as for
the manifold's coordinates. In this case the metric obtains an additional part which signature depends on the value of the gauge fields. This gauge field is a new degree of freedom in this set-up.
Another possibility is the interesting one; we introduce a non-flat tangent space of vierbein fields through the additional scalar fields with indexes related to the Lorentz group and new gauge group.
The usual viebein in this case arises as a projection of the another vierbein which we can call as gauge one.
This set-up is equivalent to the introducion of kind of a metric in the tangent space. The new scalar field is the metric there. This non-flatness allows to define the usual metric and it's signature in terms of the scalar fields which values, in turn, will depend on the values of induced gauge fields. For both cases, the usual metric can be formulated in a non-perturbative and in a perturbative manner. 
In the non-perturbative framework the manifold's metric will be defined fully in terms of the gauge fields involved non-linearly through some 4D non-linear sigma model, i.e. the action for these fields will depend on the metric which in turn is defined in terms of the gauge fields. For the perturbative case, the gauge fields provide a fluctuation of the metric with undefined signature above some fixed background. In this case the action for the gauge fields can be considered in the flat space-time in the first approximation.

  There are interesting additional problems which we do not consider in the article but which may have relation with the proposed idea. 
The complexification of the coordinates leads to the eight dimensional manifold, in this context a generalization of the approach can be achieved, for example, by the introduction of the
coordinates considered as
p-adic number field on the manifold, see \cite{Volovich}. This construction can lead to the  manifold with dimension larger than 4D dimensions, in this case more that one small parameter can be introduced. The zero values of all parameters will lead to usual real metric in this case as well, otherwise some complicated variant of the proposed framework will be obtained. 
Therefore, the metric and its signature will be valued in the field of p-adic numbers instead of complex ones.
Another face of the complexification is a similarity of the introduced phases to the 
"fast" variables of t'Hooft, \cite{Hooft}, introduced in his generalization of quantum mechanics. Formally speaking, the "fast" variables are fields in the 't Hooft approach, their counterpart in the given framework are gauge fields. Therefore, in general, it is interesting to understand the consequences of the manifold's complexification on the formulation of quantum mechanic approaches.

	The paper is organized as follows. In the next section we discuss basic ideas of a definition of the complex coordinates and metric for a complex manifold. In the Section \ref{GSym}
we consider a simplest variant of the Einstein-Cartan gravity for the complex manifold with complex coordinates, whereas in the Section \ref{LSym} we investigate the gravity for the coordinates complexified by the gauge fields. In the Section \ref{GauV} and Section \ref{NonV} vierbein based approaches to the problem are considered, firstly a investigation of the additoinal gauge symmetry for a new vierbein field and further a construction of a non-flat tangential space for the Einstein-Cartan gravity. The last section is a Conclusion of the paper.

\section{Complex metrics for a complex manifold}\label{CompM} 


  In order to clarify the ideas of the framework we, first of all, consider the following simple construction.
Let's define a complex manifold on the base of the usual real four-dimensional manifold by simple complexification of it's coordinates:
\beq\label{CM1}
p\,=\,(p^{1},\dots ,\,p^{n})\rightarrow\,z\,=\,(z^1,\dots ,\, z^{n})\,=\,(p^1 e^{\imath \phi_1},\cdots ,\, p^{n} e^{\imath \phi_n})\,.
\eeq
Defining the tangential vector fields in each $z$ of the complex and each $p$ of the real manifolds
\beq\label{CM2}
X_{z}\,\in\,T_{z}^{\mathbb{C}}\,,\,\,\,X_{p}\,\in\,T_{p}^{\mathbb{R}}\,
\eeq
we observe that the fields are connected as 
\beq\label{CM22}
X_{z}\,=\,M\,X_{p}\,
\eeq 
with $M$ as $U(4)$ diagonal matrix, see Appendix \ref{AppA} example.
Using the usual definition metric for the real manifold
\beq\label{CM21}
\g(X_{p_1},Y_{p_2})\,=\,\g((x^1,\dots ,\, x^{n}),\,(y^1,\dots ,\, y^{n} )\,)\,=\,
\g_{i j}\,x^{i}\,y^{j}\,,
\eeq
we correspondingly define the quadratic complex form on the complex manifold as
\beq\label{CM3}
\g(X_{z_1},Y_{z_2})\,=\,\g((x^1 e^{\imath \phi_{x_1}},\cdots ,\, x^{n} e^{\imath \phi_{x_n}}),\,(y^1 e^{\imath \phi_{y_1}},\cdots ,\, y^{n} e^{\imath \phi_{y_n}})\,)\,=\,
e^{\imath (\phi_{i} + \phi_{j})}\,\g_{i j}\,x^{i}\,y^{j}\,.
\eeq
Now we introduce the following complex metric in a local coordinate basis defining it as
\beq\label{CM4}
\g\,=\,e^{\imath a_{\phi} (\phi_{i} + \phi_{j})}\,\g_{i j}\,dx^{i}\otimes dx^{j}\,+\,\cdots
\eeq
with the $\g_{i j}$ as a metric field of the real manifold and $a_{\phi}$ as some  parameter. The additional parts of the metric are proportional to the new parameter $a_{\phi}$,
redefining the angles in this expressions for the usual complex coordinates we obtain:
\beq\label{CM7}
\phi\,\rightarrow\,a_{\phi}\,\phi\,.
\eeq
In order to stay in the situation with four dimensional classical world we need to push all the effects of the additional dimensions to the areas of some special
regimes. In the formalism it means that the parameter must be extremely small for the case of the classical world. We define correspondingly the following dimensionless parameter:
\beq\label{CM701}
a_{\phi}\,=\,\frac{l_{0}}{R_{0}}\,
\eeq
with $l_{0}$ as Planck length and $R_{0}$ as curvature of the manifold, i.e. the parameter proposed is extremely small indeed in the present physical reality. It's smallness has two purposes, first of all, 
the real metric appears in the model as the first term of the expansion of the complex one with respect to $a_{\phi}$. The limit
\beq\label{CM8}
a_{\phi}\,\rightarrow\,0\,
\eeq
provides the expansion with the usual metric as a leading contribution term. The second important role of this parameter, as mentioned above, is that it's smallness allows to decouple the additional metric's components
in the corresponding expression \eq{CM4}.
Namely, for a general complex metric in eight dimensional space we have  $g_{\phi x}\,\propto\,a_{\phi}$ that provides $a_{\phi}^2$ order contribution in the corresponding gravity action. Therefore, preserving 
everywhere $a_{\phi}$ order, i.e. with precision linear with respect to parameter, we can limit calculations by the four dimensional metric of the real Riemann manifold modified in correspondence to \eq{CM4} prescription. 
Due the smallness of the value of the  $a_{\phi}$ parameter we note also that the corrections related to the complexifications can contribute only at some extremal conditions. We assume that it can be important at the level of Planck length, in quantum gravity consideration and in the situation of the extremely strong gravity appearance. 
This smallness, as well, provides a simple rule for the use of diffeomorphisms to change the coordinates. For the global gauge symmetry we require that
\beq\label{CM8111} 
\frac{\D z^{\mu}}{\D p^{\nu}}\,\rightarrow\,e^{\imath a_{\phi} (\phi_{z} - \phi_{p})}\,\frac{\D x^{\mu}}{\D y^{\nu}}\,
\eeq
equivalent to the firstly done diffeomorphism transform and further complexification:
\beq\label{CM8111} 
\frac{\D x^{\mu}}{\D y^{\nu}}\,\rightarrow\,e^{-\imath  a_{\phi} (\phi_{z} - \phi_{p})}\,\frac{\D z^{\mu}}{\D p^{\nu}}\,.
\eeq
Of course in the case when $a_{\phi}\,\rightarrow\,1$ we will need to account all eight coordinates and in this case the commutation between the complexification and real diffeomorphism transformations may not work. We do not discuss this case in the article.   
The inverse metric field is defined correspondingly, in the dual basis it reads as
\beq\label{CM5}
\g^{-1}\,=\,e^{-\imath a_{\phi} (\phi_{i} + \phi_{j})}\,\g^{i j}\,\e_{i}\otimes \e_{j}\,
\eeq
where
\beq\label{CM6}
\g_{i j}\,\g^{j k}\,=\,\delta^{k}_{i}
\eeq
for the real manifold.

 We note, that we can obtain an additional example of the complex metric  
if we consider the angles as some internal parameters related to the additional symmetry of the covariant and contravariant bases of the real manifold:
\beq\label{CM81}
\e^{i}\,\rightarrow\,e^{\imath a_{\phi} \phi_{i}}\,\e^{i}\,,\,\,\,\,\e_{i}\,\rightarrow\,e^{-\imath a_{\phi} \phi_{i}}\,\e_{i}\,,
\eeq
i.e. the complexification of the basic vectors leads to the same results as complex coordinates in \eq{CM1} and \eq{CM3}.
In the following we will use the Einstein-Cartan formulation of the gravity. Therefore, considering a four dimensional real Riemann manifold and introducing the real Lorentz vierbein (tetrad) $\e^{\,a}_{\mu}$ as usual
\beq\label{CM9}
g_{\,\mu \nu}\,=\,\eta_{a b}\,\e^{\,a}_{\,\mu}\,\e^{\,b}_{\,\nu}\,.
\eeq
we can consider the corresponding complex metric defined as following:
\beq\label{CM10}
g_{\,\mu \nu}\,=\,\eta_{a b}\,e^{\imath a_{\phi} (\phi_{a} + \phi_{b}) }\,\e^{\,a}_{\,\mu}\,\e^{\,b}_{\,\nu}\,.
\eeq
This metric can be obtained from the real one by the complexification of the tetrad:
\beq\label{CM11}
\e^{\,a}_{\,\mu}\,\rightarrow\,e^{\imath a_{\phi} \phi_{a}}\,\e^{\,a}_{\,\mu}
\eeq
and 
\beq\label{CM12}
\E_{\,a}^{\,\mu}\,\rightarrow\,e^{-\imath a_{\phi} \phi_{a}}\,\E_{\,a}^{\,\mu}
\eeq
for the inverse vierbein. This construction can be considered as a particular example of the non-flat tangent space which we will discuss further.

 It is important to notice, that the \eq{CM4} expression can be considered as an approximate one and metric's complexification in this formulation arises as consequence of the decoupling of the corresponding coordinates. 
The
\eq{CM81}-\eq{CM12} construction, in contrast to that, is precise and the angles arise there due the introduced additional symmetries related to vierbein's complexification. Therefore, in the first case we consider a
complex manifold with real functions which depend on the complex coordinates, there is a integration over the angles in the action. In the second case, correspondingly, we have a real manifold with angles as parameters of corresponding independent $U(1)$ symmetry groups for each real coordinate with action invariant with respect to the symmetries, there is still only four real coordinates to integrate. Also,
whereas in the first case we need to consider small $a_{\phi}$ parameter in the problem because so far we have no observable complex manifolds, in the second case case we can take $a_{\phi}\,=\,1$ of course.
Consequently, this formulation of the approach will lead to the variant of the framework with
complex metric which was defined and discussed in \cite{Moffat} for example (see also \cite{Witten,Turok}). We do not consider this case further.

\section{Einstein-Cartan action for the complex metric in the complex manifold}\label{GSym} 

 In order to derive the analog of the Einstein-Cartan action for the complex metric introduced above, we, first of all, consider the transformation of the vector with Lorentz index projected with the 
help of the new wierbein:
\beq\label{GSym1}
\delta\,\tilde{\e}^{a}=\delta \Le e^{\imath a_{\phi} \phi_{a}}\,\e^{a} \Ra\,=\,e^{\imath a_{\phi} \phi_{a}}\,\delta\, \e^{a}=
e^{\imath a_{\phi} \phi_{a}}\,\omega^{a}\,_{b}\,e^{b} =
\Le e^{\imath a_{\phi} \phi_{a}}\,\omega^{a}\,_{b}\,e^{-\imath a_{\phi} \phi_{b}} \Ra\,\Le e^{\imath a_{\phi} \phi_{b}}\,\e^{b}\Ra=\tilde{\omega}^{a}\,_{b}\,\tilde{\e}^{b}\,,
\eeq
we see that the expression is invariant in respect to the internal symmetry transformation of the covariant and contravariant Lorentz indices performed in correspondence to the 
\eq{CM11}-\eq{CM12} definitions. In the following, denoting the complex vierbein as the usual one, we will remember that the Lorentz indices allows to rotate the corresponding objects
in correspondence to this symmetry  without mixing with the Lorentz transformations.

  Now, as mentioned above, we have to distinguish between the cases when we consider a complex manifold or we introduce the 
additional symmetry in the problem related to the $U(1)$ global gauge symmetry for each Lorentz index. There is the following form of Palatini action we have for the first case:
\beq\label{GSym2}
S\,=\,C\,\frac{m_{P}^{2}}{2}\,
\int\,d^4 z\,\varepsilon^{\mu \nu \rho \sigma}\,\varepsilon_{a b c d}\,e_{\rho}^{c}\,e_{\sigma}^{d}\,\Le D_{\mu} \om_{\nu}^{a b} \Ra\,
\eeq
with $C$ as some normalizations constant, $\D_{z}$ in the covariant derivative and $z$ as complex coordinates, see the use of the complex coordinates in the formulation of the Quantum Mechanics 
in \cite{Witten} for example\footnote{Discussions about complex path integral trajectories for Lorentz path integrals can be found in \cite{Turok}.}.
Our next step is an assumption that the integration functions are analytical in the whole region of interest, except, perhaps, some 
extreme boundary points.
Consequently, to first approximation, we can choose the integration paths for each variable $z_{\mu}$ taking $x_{\mu}\,\in\,[-\infty, \infty]$ at some fixed constant $\phi_{\mu 0}$ angles.
Therefore, assuming a smallness of $a_{\phi}$ parameter\footnote{We note here, that due the rotation of the manifold's coordinates, the direction of the rotation is important. Namely, taking 
$e^{-\imath a_{\phi} \phi}$ (conjugated) definition of the complex coordinate, we will have to change the integration limits as $0\,\rightarrow\,0$ and $2\pi\,\rightarrow\,-2\pi$  that will lead to the invariance of the \eq{GSym3} expression with respect to the definition of the $z$ coordinate.}
we write the action till the requested precision order as following:
\beqar\label{GSym3}
S\,&\approx &\,C\,e^{\imath a_{\phi} \sum_{\mu=0}^{3} \phi_{\mu 0}}\,\frac{m_{P}^{2}}{2}\,
\int\,d^4 x\,\varepsilon^{\alpha \beta \rho \sigma}\,\varepsilon_{a b c d}\,e_{\rho}^{c}\,e_{\sigma}^{d}\,\Le D_{\alpha} \om_{\beta}^{a b} \Ra\,+\,
\nonumber \\
&+&\,
\imath\,a_{\phi}\,C\,\frac{m_{P}^{2}}{2}\,\sum_{\mu,\,\nu\neq \mu,\,}^{3}\,e^{\imath a_{\phi}\sum_{\nu}^{3} \phi_{\nu 0}}\,\int\,x^{\mu}_{0}\,d \phi_{\mu}\,d^{3} x\,
e^{\imath a_{\phi} \phi_{\mu}}\,\varepsilon^{\alpha \beta \rho \sigma}\,\varepsilon_{a b c d}\,e_{\rho}^{c}\,e_{\sigma}^{d}\,\Le D_{\alpha} \om_{\beta}^{a b} \Ra\,.
\eeqar
The condition when the usual Einstein-Cartan formalism is reproduced in the first order approximation is the following one:
\beq\label{GSym6}
\,C\,e^{\imath a_{\phi} \sum_{\mu=0}^{3} \phi_{\mu 0}}\,=\,1\,.
\eeq
There are some arbitrary constant angles $\phi_{i\,0}$ introduced here and this condition can be considered as definition of the $C$ constant as well. 
We also note, that the expression under the integration is function of $z$
and in general it must be expanded as well in order to provide all $a_{\phi}$ order corrections to the real action.

  The interesting consequence of the form of \eq{GSym3} action is that it does not define any preferable direction of time or preferable value of the signature.
Indeed, let's choose the special coordinate system, x-system, with
\beq\label{GSym7}
\phi_{\mu\,0}\,=\,0\,,\,\,\,\,\mu\,=\,0 \ldots 3\,;\,\,\,\,C\,=\,1\,.
\eeq
In the same way we can choose any other angles such that
\beq\label{GSym8}
\sum_{\mu=0}^{3} \phi_{\mu\,0}\,\neq\,0\,,\,\,\,\,\phi_{\mu 0}\,\neq\,0\,,\,\,\,\,C\,=\,1\,,
\eeq
the angles define a new coordinate system different from the special one. Namely, for the non-zero $\phi_{i\,0}$ there are new coordinates 
\beq\label{GSym9}
y^{\mu}\,=\,x^{\mu}\,e^{\imath\,a_{\phi}\,\phi_{\mu 0}}\,.
\eeq
In terms of the new coordinates the action acquires the following form:
\beqar\label{GSym10}
S\,&= &\,\frac{m_{P}^{2}}{2}\,
\int\,d^4 y\,\varepsilon^{\alpha \beta \rho \sigma}\,\varepsilon_{a b c d}\,e_{\rho}^{c}\,e_{\sigma}^{d}\,\Le D_{\alpha} \om_{\beta}^{a b} \Ra\,+\,
\nonumber \\
&+&\,
\imath\,a_{\phi}\,\frac{m_{P}^{2}}{2}\,\sum_{\mu=0}^{3}\,\int\,y^{\mu}_{0}\,d \chi_{\mu}\,d^{3} y\,
e^{\imath a_{\phi} \chi_{\mu} }\,\varepsilon^{\alpha \beta \rho \sigma}\,\varepsilon_{a b c d}\,e_{\rho}^{c}\,e_{\sigma}^{d}\,\Le D_{\alpha} \om_{\beta}^{a b} \Ra\,
\eeqar
where
\beq\label{GSym5}
\phi_{\mu}\,=\,\phi_{\mu 0}\,+\,\chi_{\mu}\,.
\eeq
With redefinition of the arguments of the integral functions
performed in \eq{GSym9}  and subsequent deformation of the integration contours, the form of the redefined action is the same as \eq{GSym3} with \eq{GSym7} values of the angles. The only different contribution into the action, therefore, can come from the end points of the integration over real $y^{\mu}$ which acquire complex phases in the case 
of \eq{GSym9} variables change. We assume that these contributions are zero.
As a result of the complexification of the manifold and its symmetry we have
an infinite number of the equivalent directions of time and foliations of the spatial coordinates\footnote{This statement can be understood in terms of any evolution equations,
the equations will have the same form for any redefined $x$ coordinate.} for the given
value of a signature.
An another interesting consequence of the model is that the \eq{GSym10} action 
quite naturally acquires a small term additional to the leading one. This term can be considered as a type of the cosmological constant in the action in the framework of the perturbative scheme based on the 
$a_{\phi}$ smallness. In general it means that the term must be finite after the integration over $\chi_{\mu}$ angle at some $y_{\mu}\,\rightarrow\,y_{\mu 0}$ limits taken in the corresponding contour integral.

 Now, expanding the vierbein field in the new perturbative scheme as
\beq\label{GSym1001}
e_{\nu}^{c}\,=\,e_{\nu\,0}^{c}\,+\,e_{\nu\,1}^{c}\,
\eeq
and taking a variation of the \eq{GSym3} action with respect to $\omega$ connections we will obtain:
\beq\label{GSym11}
\D_{[\mu}\,e_{\nu]\,1}^{c}\,=\,-\,\imath\,a_{\phi}\,\sum_{\rho\,=\,0}^{3}\,x^{\rho}\,\delta(x^{\rho}\,-\,x^{\rho}_{0})\,\int\,d \phi_{\rho}\,e^{\imath a_{\phi} \phi_{\rho} }\,\D_{[\mu}\,e_{\nu]\,0}^{c}\,
\eeq
or, equivalently
\beq\label{GSym12}
\int\,d^4 x\,\Le \D_{\mu}\,e_{\nu\,1}^{c}\,+\,
\imath\,a_{\phi}\,\sum_{\rho\,=\,0}^{3}\,x^{\rho}\,\delta(x^{\rho}\,-\,x^{\rho}_{0})\,\int\,d \phi_{\rho}\,e^{\imath a_{\phi} \phi_{\rho} }\,\D_{\mu}\,e_{\nu\,0}^{c}\Ra\,=\,0\,.
\eeq
Providing some initial value of $e_{\nu\,1}^{c}(x)$ at $x^{\mu}_{0}$
we can write the solution of \eq{GSym11} as
\beq\label{GSym13}
e_{\nu\,1}^{c}\,=\,e_{\nu\,1}^{c}(x_{0}^{\mu})\,-\,\imath\,a_{\phi}\,\sum_{\rho\,=\,0}^{3}\,\int_{x_{0}^{\mu}}^{x^{\mu}}\,dx^{\mu}\,
\Le x^{\rho}\,\delta(x^{\rho}\,-\,x^{\rho}_{0})\Ra\,\int\,d \phi_{\rho}\,e^{\imath a_{\phi} \phi_{\rho} }\,\D_{\mu}\,e_{\nu\,0}^{c}\,.
\eeq
This additional virbein's part provides a correction to the metric through \eq{CM9} definition. It can be of any signature depending on the value of the integral in \eq{GSym13} r.h.s..

\section{Complexification of the manifold through gauge fields}\label{LSym} 

 If we consider the simplest generalization of the \eq{CM1}  complexification through the replacement of the $\phi_{\mu}$ angles by $\phi_{\mu}(p)$ functions 
then we immediately realize that this construction does not work. Namely, in this setup there is no self-consistent definition
of the coordinates and corresponding functions in the integrals. 
Therefore, in order to discuss the case of the local complexification of the real manifold, we will consider 
the following model. We introduce a set of real coordinates $x^{\mu}$ and determine the new coordinates of the manifold as transform of $x$ :
\beqar\label{LSym1}
z^{\alpha}\,& = &\,\,M^{\,\alpha}\,_{\mu}(x)\,x^{\mu} \nonumber \\
z_{\alpha}\,& = &\,\,M_{\,\alpha}\,^{\mu}(x)\,x_{\mu} 
\eeqar
with new vierbein like gauge fields  $M$,  where the new indices are transforming in correspondence to some group $G$ .
Accordingly, we will consider the integrals as taken over the Riemann $x^{\mu}$ with functions defined as depending on  $z^{\alpha}$ similarly to done before.
Introducing an another form of the gauge fields, suitable for the perturbative calculations\footnote{We note that there is a difference between gauge actions written in terms of $M$ and $A$ fields.}, we can define the complex coordinates in this case as
\beq\label{LSym2}
z^{\alpha}\,=\,\Le \delta^{\alpha}_{\mu}\,+\,\imath\,a_{\phi}\,A^{\alpha}\,_{\mu}(x)\Ra\,x^{\mu}\,
\eeq
with $A$ as some gauge field related to the $G$ group.
The coupling constant $a_{\phi}$ is small again and determines a measure of the complexification. In this setup
it plays a role of the coupling constant of the new gauge group $G$.

  The new gravity action of the model preserves the form of the \eq{GSym2} action and  we have:
\beq\label{LSym3}
S\,=\,C\,\frac{m_{P}^{2}}{2}\,
\int\,d^4 x\,\varepsilon^{\mu \nu \rho \sigma}\,\varepsilon_{a b c d}\,e_{\rho}^{c}\,e_{\sigma}^{d}\,\Le D_{\mu} \om_{\nu}^{a b}\Ra \,
\eeq
with functions in the integral depending on $z$ coordinates. 
There is, correspondingly, an additional part in the action which corresponds to the new introduced field which we consider as a gauge one.
Using an usual determination of the  field's strength of the new gauge field $A$ through the covariant derivative
\beq\label{LSym4}
[D_{G\mu}\,D_{G\nu}]\,=\,-\,\imath\,a_{\phi}\,G_{\mu \nu}^{\alpha}\,t^{\alpha}
\eeq
for some representation of the $G(N)$ 
\beq\label{LSym5}
[t^{\alpha},\,t^{\beta}]\,=\,\imath\,f^{\alpha \beta \gamma}\,t^{\gamma}\,,
\eeq
we define this action as
\beq\label{LSym6}
S_{A}\,=\,\kappa\,\int\,d^4 x\,e\,tr \left[ G_{\mu \nu}\,G_{\mu_1 \nu_1} \right]\,e^{\mu}_{a}\,e^{\nu}_{b}\,e^{\mu_1}_{a_1}\,e^{\nu_1}_{b_1}\,
{\cal F}^{a b;a_1 b_1}\,
\eeq
with mostly general
\beq\label{LSym7}
{\cal F}^{a b;a_1 b_1}\,=\,\kappa_{1}\,\eta^{a a_1}\,\eta^{b b_1}\,+\,\kappa_{2}\,\eta^{a b_1}\,\eta^{b a_1}\,+\,
\kappa_{3}\,\eta^{a b}\,\eta^{a_1 b_1}\,
\eeq
with $\eta$ as Lorentz metric of the flat space.
The action is similar, for example, to the QCD action in the curved space time. The interaction
between these two parts of the action can be written in terms of the expansion of  \eq{LSym3} functional with respect to the complex part of the
$z$ coordinates.

 Our next step is an introduction of the non-trivial $A$ gauge fields in the action.
The idea is the following.
We can introduce in the action a following additional term:
\beq\label{VC121}
S_{ind}\,= \,\frac{m_{P}^{2}}{2}\,\sum_{i}\,\int\,d^{4}x\,T_{\mu i}(A)\,\mathscr{ A}^{\,\mu}_{\,i}\,
\eeq
which we call an induced part of the action in correspondence to the definition of \cite{EffA,EffA1}. The purpose of this part of the action is to introduce in the equations of motion
the classical values of the $A$ gauge field, denoted as $\mathscr{ A}$, which satisfy some boundary conditions at the edges of time interval.
Namely, we define the boundary conditions at some different limits  of $t$ coordinate:
\beqar\label{VC131}
&\,&\,\delta_{\omega}\, T_{\mu i}(A)\,=\,J_{\mu i}(A)\,\delta A \,\nonumber \\
&\,&\, J_{\mu i}(t_{j})\,\rightarrow\,0\,,\,\,\,\,t\,\rightarrow\,\,t_{0 j\neq i}\,\label{VC14} \\
&\,&\,A_{cl \mu }\,=\,\sum_{i}\,\mathscr{ A}_{\,\mu\, i}\,
\eeqar
with the last equation fixed by the structure of \eq{VC121} term and obtained from the usual equations of motion:
\beq\label{VC151}
\delta_{A}\,\Le \,S_{A}\,+\,S_{ind}\,\Ra\,=\,0\,.
\eeq
In general, the effective currents $T_{\mu i}$ can be consequently reconstructed by requests of the gauge invariance of the induced part of the action,
with details that can be found in  \cite{EffA,EffA1}.
Fixing the boundary conditions, 
i.e. fixing the values of $A_{cl i}$ at the edges of the overall time interval for example, we will obtain 
an  action with some space-time foam  above the  background space-time of a fixed signature. The example of this construction is given in the Appendix \ref{AppB}. Taking only one boundary field
from \eq{VC21} expression for example, we will have 
\beq\label{VC161}
A_{cl \mu }\,=\,\mathscr{ A}_{\,\mu}.
\eeq
Now, writing the equations of motion for $\omega$ connections
\beq\label{VC171}
D_{[\mu}\,e_{\nu]}^{c}\,=\,0\,,\,\,\,\,\frac{\D}{\D x^\mu}\,=\,\frac{\D z^{\nu}}{\D x^\mu}\,\frac{\D}{\D z^\nu}\,,
\eeq
and expanding the vierbein in a perturbative scheme related with the parameter $a_{\phi}$ as 
\beq\label{VC181}
e_{\nu}^{c}\,=\,e_{\nu\,0}^{c}\,+\,e_{\nu\,1}^{c}\,,
\eeq
we will obtain then 
\beq\label{VC191}
\D_{[\mu}\,e_{\nu]\,1}^{c}\,=\,-\,\imath\,a_{\phi}\,\D_{\,[\mu}\Le \mathscr{ A}_{\,\rho}^{\alpha} x^{\rho} \Ra\,\D_{\alpha} e_{\nu]\,0}^{c}
\eeq
with solution
\beq\label{VC201}
e_{\nu\,1}^{c}\,=\,e_{\nu\,1}^{c}(x^{\mu}_{0})\,-\,\imath\,a_{\phi}\,\int_{x_{0}^{\mu}}^{x^{\mu}}\,dx^{\mu}\,\D_{\,\mu}\Le \mathscr{ A}_{\,\rho}^{\alpha} x^{\rho} \Ra\,\D_{\alpha} e_{\nu\,0}^{c}
\eeq
for the additional vierbein's part. We see, that the vierbein acquires a correction which structure is defined by the value of the boundary $\mathscr{ A}_{\,\mu}$ fields.
The corresponding \eq{CM9} metric obtains an additional part as well and the signature of this metric's correction is depending on the $\mathscr{ A}_{\,\mu}$ fields.
This situation, as we will see further,  will realized in other models with gauge field involved.

 We note, that the \eq{VC161} $A_{cl \mu }$ field can appear in the equation as a solution of classical equations of motion which contributes mostly in the generating functional of the theory,
i.e. as semi-classical solution of the theory. This possibility is a purely dynamical one and requires an analysis of the classical dynamics of whole system under consideration. For example, it is not clear, how dynamically solutions of following type 
\beq\label{VC1611}
A_{cl \mu }\,=\,\mathscr{ A}_{1\,\mu}\,+\,\mathscr{ A}_{2\,\mu}\,
\eeq
with two or more different classical $\mathscr{ A}_{i\,\mu}$ fields will arise for a connected manifold. Further we discuss this mechanism only in the Conclusion of the paper.

 It is interesting to note also, that assuming the existence of a the transform opposite to \eq{LSym1} 
\beqar\label{VC211}
x^{\mu}\,& = &\,\,N^{\,\mu}\,_{\alpha}(z)\,z^{\alpha} \nonumber \\
x_{\mu}\,& = &\,\,N_{\,\mu}\,^{\alpha}(z)\,z_{\alpha} 
\eeqar
we can rewrite the \eq{LSym3} action fully in terms of $z$ variable as follows
\beq\label{VC221}
S\,=\,C\,\frac{m_{P}^{2}}{2}\,
\int\,d^4 z\,\tilde{N}\,\varepsilon^{\mu \nu \rho \sigma}\,\varepsilon_{a b c d}\,e_{\rho}^{c}\,e_{\sigma}^{d}\,\Le D_{\mu} \om_{\nu}^{a b}\Ra \,
\eeq
with $\tilde{N}$ as Jacobian of the \eq{VC21} coordinates transform given by the 
\beq\label{VC231}
\tilde{N}^{\mu}\,_{\alpha}\,=\,N^{\,\mu}\,_{\alpha}\,+\,\frac{\D N^{\mu}\,_\beta}{\D z^{\alpha}}\,z^{\beta}\,
\eeq
matrix. In this case the value of the factor in front of the action in the path integral will be determined by the value of the $\tilde{N}$. 
Therefore, again, the redefinition of the $C$ factor in \eq{VC221} will lead to the equivalent actions for the different metrics with different phases for their components.

\section{Gauge symmetry for the vierbein field}\label{GauV} 

 The complexification mechanism discussed in the previous Section can be applied for the vierbein fields as well. Considering the usual vierbein use in definition of the metric
\beq\label{VC01}
g_{\,\mu \nu}\,=\,\eta_{S\,a b}\,\e^{\,a}_{\,\mu}\,\e^{\,b}_{\,\nu}\,,
\eeq
we can generalize the \eq{CM11}-\eq{CM12} definition of the complex vierbein:
\beq\label{VC02}
\e^{a}_{\mu}\,=\,M_{\mu}\,^{\,\alpha}(x)\,\e^{a}_{\alpha}\,
\eeq
or similarly to done before as
\beq\label{VC021}
\e^{a}_{\mu}\,=\,
\Le \delta^{\alpha}_{\mu}\,+\,\imath\,a_{\phi}\,A_{\mu}\,^{\alpha}(x)\Ra\,\e^{a}_{\alpha}\,.
\eeq
In these cases the \eq{VC01} metric acquires an additional part with signature which depends on the value of the fields. We have for the first metric:
\beq\label{VC031}
g_{\,\mu \nu}\,=\,\eta_{S\,a b}\,M_{\mu}\,^{\,\alpha}(x)\,M_{\nu}\,^{\,\beta}(x)\,\e^{a}_{\alpha}\,\e^{b}_{\beta}\,
\eeq
and correspondingly for the second at linear approximation with respect
to $a_{\phi}$ parameter:
\beq\label{VC03}
g_{\,\mu \nu}\,=\,\eta_{S\,a b}\,\e^{\,a}_{\,\mu}\,\e^{\,b}_{\,\nu}\,+\,\imath\,a_{\phi}\,\eta_{S\,a b}\,\Le\, A_{\mu}\,^{\alpha}(x)\,\e^{a}_{\alpha}\,\e_{\nu}^{b}\,+\,
A_{\nu}\,^{\alpha}(x)\,\e^{b}_{\alpha}\,\e_{\mu}^{a}\,\Ra\,.
\eeq
The \eq{VC031}-\eq{VC03} expressions are different in general. Whereas the \eq{VC031} metric describes a manifold with arbitrary signature which depends on the value
of $M$ matrix, the \eq{VC03} metric determines a manifold with an additional part above the background metric with given signature.
We note that this additional metric's part can be of any signature as well, it depends on the value of $A$ gauge field. In both cases,  the values of the gauge fields are determined dynamically 
through the corresponding Lagrangians.

 The Einstein-Cartan action can be easily rewritten in terms of new vierbein in this case. We require that the additional metricity condition must be satisfied:
\beq\label{VC04}
\nabla_{\mu} \Le A_{\nu}\,^{\alpha}\,\e_{\alpha}^{a} \Ra\,=\,\Le \nabla_{\Gamma\,\mu} \,A_{\nu}\,^{\alpha}\Ra\,\e_{\alpha}^{a} \,+\,
A_{\nu}\,^{\alpha}\,\Le D_{\mu}\,\e_{\alpha}^{a} \Ra\,=\,0
\eeq
with
\beqar\label{VC05}
&\,& \nabla_{\Gamma\,\mu} \,A_{\nu}\,^{\alpha}\,=\,\D_{\mu}\,A_{\nu}\,^{\alpha}\,-\Gamma^{\rho}_{\mu \nu}\,A_{\rho}\,^{\alpha} 
\\
&\,&
\label{VC06}
D_{\mu}\,\e_{\alpha}^{a}\,=\,\D_{\mu}\,\e_{\alpha}^{a}\,+\,\omega_{\mu}\,^{a}\,_{b}\,\e_{\alpha}^{b}\,,
\eeqar
here $\Gamma$ and $\omega$ are Christoffel and Lorentz connections correspondingly. We will obtain then:
\beq\label{VC07}
S\,=\,\frac{m_{P}^{2}}{2}\,
\int\,d^4 x\,
\varepsilon^{\mu \nu \rho \sigma}\,\varepsilon_{a b c d}\,M_{\rho}\,^{\alpha}\,M_{\sigma}\,^{\beta}\,
e_{\alpha}^{c}\,e_{\beta}^{d}\,\Le D_{\mu} \om_{\nu}^{a b} \Ra\,
\eeq
with obvious corresponding redefinition in terms of $A$ field. 
This part of the full action is correct for any form of the metric, the non-triviality of the construction, therefore,  is manifested through the additional gauge field. 
Introducing the gauge
field strength 
\beq\label{VC09}
[D_{G\,\mu}\,D_{G\,\nu}]\,=\,-\,\imath\,a_{\phi}\,G_{\mu \nu}\,
\eeq
for some symmetry group $G$, we define the additional part of the action as
\beq\label{VC10}
S_{A}\,=\,\kappa\,\int\,d^4 x\,\e\,M\,
tr \left[ G_{\mu \nu}\,G_{\mu_1 \nu_1} \right]\,{\cal F}^{\mu \nu;\mu_1 \nu_1}\,
\eeq
with
\beq\label{VC11}
{\cal F}^{\mu \nu;\mu_1 \nu_1}\,=\,\kappa_{1}\,g^{\mu \nu}\,g^{\mu_1 \nu_1}\,+\,\kappa_{2}\,g^{\mu \nu_1}\,g^{\mu_1 \nu}\,+\,
\kappa_{3}\,g^{\mu \mu_1}\,g^{\nu \nu_1}\,,
\eeq
and
\beq\label{VC08}
M\,=\,{\rm det}(M_{\mu}\,^{\alpha})\,,\,\,\,\,\,\e\,=\,{\rm det}(\e_{\alpha}^{a})\,,
\eeq
with matrix $M$ determined or through \eq{VC031} or either the \eq{VC03} expressions.
The metric $g^{\mu \nu}$, in turn, as well depends on the corresponding gauge field.  
The \eq{VC10} Lagrangian describes a variant of 4D non-linear gravitational sigma-model.
Correspondingly, further, we will consider only the \eq{VC03} formulation of the metric, since there is no a simply perturbative expansion in 
respect to $a_{\phi}$ for the $M$ field in the \eq{VC031} metric. Therefore, a self-consistent solution of the equations of motion  for $M$ field through \eq{VC10} is a non-trivial task.
We will consider it in an separate piblication. Concerning the \eq{VC03} metric and the $A$ gauge field, 
there is the $a_{\phi}$ parameter in the \eq{VC03} definition of the
metric so we need to know a solution for the gauge field till $a_{\phi}^{0}$ precision only and, therefore, 
for our purposes it will be enough to consider the \eq{VC10} action in the flat space-time. In this case, with the help of Appendix \ref{AppB} results, we obtain:
\beq\label{VC2101}
A^{\mu}_{\,cl}\,=\,\mathscr{ A}^{\,\mu}_{1}\,+\,\mathscr{ A}^{\,\mu}_{2}\,.
\eeq
Correspondingly, the \eq{VC03} metric will acquire an additional part determined by the $\mathscr{ A}^{\mu}_{i}$ fields: 
\beq\label{VC22}
g_{\,\mu \nu}\,=\,\eta_{S\,a b}\,\e^{\,a}_{\,\mu}\,\e^{\,b}_{\,\nu}\,+\,\imath\,a_{\phi}\,\eta_{S\,a b}\,\Le\,  A_{\mu\,cl}\,^{\alpha}(x)\,\e^{a}_{\alpha}\,\e_{\nu}^{b}\,+\,
 A_{\nu\,cl}\,^{\alpha}(x)\,\e^{b}_{\alpha}\,\e_{\mu}^{a}\,\Ra\,,
\eeq
we see that the metric's fluctuations as well as their signature are determined by the induced boundary fields of the problem.

 Considering the same approach for the \eq{VC02} $M$ fields, we will have a difference between covariant and contravariant gauge fields.
The $M$ gauge fields provide a non-flat metric initially, the field will appear in the relations between covariant and contravariant components in \eq{VC17} action therefore. It makes the problem even more non-linear. There are different vectors in the power of the \eq{VC23} ordered exponential and in the induced action. 

\section{Non-flat tangential space construction}\label{NonV}

  Geometrizing the proposed ideas, we can define the Lorentz vierbein and it's inverse as a projection of other vierbein fields:
\beqar\label{V01}
& \, &\e^{\,a}_{\,\mu}\,= \,M^{\,a \alpha}\,e_{\,\mu \alpha},\,\,\,\E_{\,a}^{\mu}\,=\,M_{\,a \alpha}\,E^{\,\mu \alpha}\,; \nonumber \\
& \, & e_{\,\mu \alpha}\,E^{\,\mu \beta}\,=\,\delta^{\beta}_{\alpha}, \,\,\,e_{\,\mu \alpha}\,E^{\,\nu \alpha}\,=\,\delta^{\nu}_{\mu}\,; \nonumber \\
& \, & e_{\,\mu \alpha}\,e^{\,\mu}_{ \beta}\,=\,M_{\,\alpha \beta},\,\,\,E^{\,\mu \alpha}\,E_{\,\mu}^{ \beta}\,=\,M^{\,\alpha \beta}\,,
\eeqar
here Greek indices $\alpha ,\,\beta$ belong to some group $G$, Latin indices denote the Lorentz transforms, the  $\mu,\,\nu$ are used as usual Riemann type indices
\footnote{An another variant of the non-flat tangent space is simply define $g_{\, \mu \nu }\,=\,M_{\alpha \beta}\,e_{\mu}^{\alpha} e_{\nu}^{\beta}$ with $M$ belonging to some extended symmetry group
with changing signature, see \cite{BonZub}.}.  
We note, that there is no general prescription to consider the dimension of the 
$G$ equal to the one of the Lorentz group,
see for example \cite{Spont}. Nevertheless, in the following we will take $\alpha,\,a\,=\,0\,\dots\,3$.
Also, unlike the previous Chapter, the $M$ field here is a scalar one, there is an additional dynamics present therefore. 
Now, 
using again the usual definition of flat metric in terms of Lorentzian vierbein
\beq\label{V02}
g_{\,\mu \nu}\,=\,\eta_{S\,a b}\,\e^{\,a}_{\,\mu}\,\e^{\,b}_{\,\nu}\,,
\eeq
with $\eta_{S\,a b}$ as a flat metric of the tangent space with some signature $S$, we rewrite it as follows: 
\beq\label{V04}
g_{\,\mu \nu}\,=\,\eta_{S\,a b}\,M^{\, a \,\alpha }\,M^{b \beta}\,e_{\,\mu \alpha}\,e_{\,\nu \beta}\,=\,
M_{S^{'}}^{\, \alpha \beta}\,e_{\,\mu \alpha}\,e_{\,\nu \beta}\,,\,\,\,\,M_{S^{'}}^{\, \alpha \beta}\,=\,M_{S^{'}}^{\,\beta \alpha }\,
\eeq
and
\beq\label{V041}
g^{\,\mu \nu}\,=\,\eta_{S}^{\,a b}\,M_{\, a \,\alpha }\,M_{b \beta}\,E^{\,\mu \alpha}\,E^{\,\nu \beta}\,=\,M_{S^{'}\, \alpha \beta}\,E^{\,\mu \alpha}\,E^{\,\nu \beta}
\eeq
with signature $S^{'}$ which can be different from $S$. Here we define 
\beq\label{V042}
M_{a \alpha}\,M^{b \alpha}\,=\,\delta^{b}_{a}\,,\,\,\,\,M_{a \alpha}\,M^{a \beta}\,=\,\delta^{\beta}_{\alpha}\,.
\eeq
The invariance of the new scalar product with respect to the new group of the symmetry is provided by the ordinary transformation rules for the new upper and lower indices
\beq\label{V17}
e_{\mu \alpha}\,=\,G_{\alpha}\,^{\beta}\,e_{\mu \beta}\,,\,\,\,\,E^{\alpha}_{\mu}\,=\,\tilde{G}^{\alpha}\,_{\beta}\,E_{\mu}^{\beta}\,,
\eeq
with $\tilde{G}$ matrix as inverse to $G$ 
\beq\label{V171}
G_{\alpha}\,^{\beta}\,\tilde{G}^{\alpha}\,_{\gamma}\,=\,\Le G\,\tilde{G}^{T} \Ra^{\beta}\,_\gamma\,=\,\Le \tilde{G}^{T}\,G \Ra^{\beta}\,_\gamma\,=\,\delta^{\beta}_{\gamma}\,.
\eeq
Both $G$ and $\tilde{G}$ matrices belong to the group of interest of course.
Correspondingly, we introduce the new covariant derivatives of the vierbein and $M$ fields with respect to the $G$ symmetry group:
\beq\label{V181}
D_{G\,\mu}\,e_{\nu\,\alpha}\,=\,\D_{\mu}\,e_{\nu\,\alpha}\,-\,\Omega_{\mu \alpha}^{\beta}\,e_{\nu\,\beta}
\eeq
and
\beq\label{V182}
D_{G\,\mu}\,M^{a\,\alpha}\,=\,\D_{\mu}\,M^{a\,\alpha}\,+\,\Omega_{\mu\beta}^{\alpha}\,M^{a\,\beta}\,.
\eeq
Here
\beq\label{V1821}
\Omega_{\mu\beta}^{\alpha}\,=\,\imath\,a_{\phi}\,\Omega_{\mu}^{a}\,\Le t_{a}\Ra^{\alpha}\,_{\beta}\,
\eeq
is a new gauge field additional to the usual connection field in the corresponding covariant derivative of the Einstein-Cartan gravity Lagrangian.

  The form of the Einstein-Cartan gravity action is changing trivially in this version of the formalism. We require the metricity property of the new vierbein in respect to the full covariant derivative
\beq\label{Tn4}
\nabla\,E_{a}^{\mu}\,=\,\nabla \Le M_{a \alpha}\,E^{\mu \alpha} \Ra\,=\,0\,
\eeq
and obtain
\beq\label{Tn5}
S_{\omega}\,=\,-\,m_{p}^{2}\,\int\,d^4 x\,\sqrt{-g}\,E^{\mu}_{a}\,E^{\nu}_{b}\,{\cal R}_{\mu \nu}\,^{a b}\,\rightarrow\,
S\,=\,-\,m_{p}^{2}\,\int\,d^4 x\,e\,M\,\Le M_{a \alpha}\,E^{\mu \alpha} \Ra\,\Le M_{b \beta} E^{\nu \beta}\Ra \,{\cal R}_{\mu \nu}\,^{a b}\,
\eeq
in correspondence to the \eq{V04} definition,
here 
\beq\label{Tn2}
M\,=\,{\rm det}\Le M^{\alpha a}\Ra\,.
\eeq
The invariant action for the $M$ field we can write as the usual action for a scalar field:
\beq\label{Tn1}
S_{M}\,=\,
\int\, d^{4} x\,\,e\,M\,
e^{\,\mu}_{ \alpha}\,e^{\,\nu}_{ \beta}\,\Le D_{G\,\mu} M \Ra^{\,c \gamma}\,\Le D_{G\,\nu} M \Ra_{c}\,^{\rho}\,{\cal F}^{\alpha \beta}_{\gamma \rho}\,
\eeq
with
\beq\label{Tn3}
{\cal F}^{\alpha \beta}_{\gamma \rho}\,=\,\alpha_{1}\,M^{\alpha \beta}\,M_{\gamma \rho}\,+\,
\alpha_{2}\,M^{\alpha}\,_{\gamma}\,M^{\beta}\,_{\rho}\,+\,\alpha_{3}\,M^{\alpha}\,_{\rho}\,M^{\beta}\,_{\gamma}\,.
\eeq
A new, in comparison to the previous section, term of the action is a free action term of the $\Omega$ gauge field. Determining the field's strength of the new gauge field
\beq\label{Tn6}
[D_{G\,\mu}\,D_{G\,\nu}]\,=\,-\,G_{\mu \nu}\,
\eeq
we define this action as
\beq\label{Tn7}
S_{\Omega}\,=\,\kappa\,\int\,d^4 x\,e\,M\,e^{\mu}_{\alpha}\,e^{\nu}_{\beta}\,e^{\mu_1}_{\alpha_1}\,e^{\nu_1}_{\beta_1}\,
tr \left[ G_{\mu \nu}\,G_{\mu_1 \nu_1} \right]\,{\cal F}^{\alpha \beta;\alpha_1 \beta_1}\,
\eeq
with
\beq\label{Tn8}
{\cal F}^{\alpha \beta;\alpha_1 \beta_1}\,=\,\kappa_{1}\,M^{\alpha \alpha_1}\,M^{\beta \beta_1}\,+\,\kappa_{2}\,M^{\alpha \beta_1}\,M^{\beta \alpha_1}\,+\,
\kappa_{3}\,M^{\alpha \beta}\,M^{\alpha_1 \beta_1}\,.
\eeq
The action is similar, for example, to the QCD action in the curved space time, we do not consider a torsion and a cosmological constant terms in the action.

 The dynamics of the theory given by \eq{Tn1} and \eq{Tn7} actions is non-linear and pretty complicated. Therefore, postponing the precise derivation for an additional publication, we can understand 
a dynamical signature in this variant of the theory by the following simple observations. First of all, we assume that for the \eq{Tn7} action there exists a classical solution for the gauge fields provided by the
mechanism described in the Appendix \ref{AppB}. We will have then:
\beq\label{Tn9}
\Omega_{\mu \beta\,cl}^{\alpha}\,=\,\mathscr{A}_{\mu \beta}^{\alpha}\,,
\eeq
where the $\mathscr{A}_{\mu}$ fields, again, are known and satisfy some boundary conditions.
This result, of course, is a consequence of the constant form of the vierbeins fields $e$ and $M$ in \eq{Tn7}, we take these fields as normalized to the delta functions with respect to the
corresponding indices in the first approximation. In this case the \eq{Tn7} will acquire the form of \eq{VC17} action. Secondly, the classical solution of the \eq{Tn1} action is provided by the following equation:
\beq\label{Tn10}
D_{G\,\mu}\,M^{a\,\alpha}\,=\,0\,
\eeq
the solution can be written:
\beq\label{Tn11}
M^{a\,\alpha}(x)\,=\,M^{a\,\beta}_{0}\,\Le P\,e^{-\imath\,a_{\phi}\,\int_{-\infty}^{x}\,d z^{\mu}\,\mathscr{A}_{\mu}(z)} \Ra_{\beta}^{\alpha}\,.
\eeq 
Again, the signature of the \eq{V04} metric is determined by the values of the boundary gauge fields $\mathscr{A}_{\mu}$ 
\beq\label{Tn12}
g_{\,\mu \nu}\,=\,\eta_{S\,a b}\,M^{a\,\alpha_1}_{0}\,M^{b\,\alpha_2}_{0}\,
\Le P\,e^{-\imath\,a_{\phi}\,\int_{-\infty}^{x}\,d z^{\mu}\,\mathscr{A}_{\mu}(z)} \Ra_{\alpha_1}^{\alpha}\,
\Le P\,e^{-\imath\,a_{\phi}\,\int_{-\infty}^{x}\,d z^{\mu}\,\mathscr{A}_{\mu}(z)} \Ra_{\alpha_2}^{\beta}\,
e_{\,\mu \alpha}\,e_{\,\nu \beta}\,
\eeq
and in principle can be arbitrary, see Appendinx \ref{AppA} 
for the similar simple example. 



 \section{Conclusion}

  In this note we consider approaches  where the signature of the metric is undefined and takes values in the field of complex numbers.
We discuss a few possibilities for the definition of this type of metric, with signs of it's components are not fixed in general.  	
The change of the signature was widely discussed in the literature, see for example \cite{MisWh,Hawk1,Sakh,Ander,Gero,Sork, Strom}, for a description of the transition form Lorentzian to Euclidean
manifold types in the quantum gravity and quantum cosmology.
Nevertheless, mostly, this transition was introduced by the time's coordinate Wick rotation.  We, instead, propose the formalism where 
the domain of the metric's signature is expanded. The metric in the proposed approaches is a dynamical object with signature determined by the complexification of the space-time manifold or by new gauge fields.	Therefore, the signature can be changed  smoothly between any predefined signatures, Lorentzian and Euclidean for example, with the help of the gauge fields. For example,
introducing the Euclidean $\mathscr{ A}$ and Lorentzian $\mathscr{ B}$ gauge fields we can calculate an effective action defined in respect to these fields:
\beq\label{Con1}
\Gamma(\mathscr{ A},\mathscr{ B})\,=\,\sum\,\mathscr{ A}_{1}\cdots\mathscr{ A}_{i}\,C^{1\cdots i;1\cdots k}\,\mathscr{ B}_{1}\cdots\mathscr{ B}_{k}\,.
\eeq
The corresponding generating functional
\beq\label{Con2}
Z(\mathscr{ A},\mathscr{ B})\,=\,Z_{0}^{-1}\,\int\,D \Phi\,e^{\imath\,\Gamma(\mathscr{ A},\mathscr{ B})}\,
\eeq
with  $\Phi$ as  all other fields in the framework,
will determine a transitional amplitude (S-matrix) between the manifolds with different signature. A calculation of those S-matrix elements we reserve for the future research.
Another possibility for such S-matrix construction is an appearance of the $\mathscr{ A}_{cl}$  field as 
a semi-classical solution,  i.e. saddle point, of the equations of motion for the total action in the generating functional, see \cite{Witten,Turok} for the corresponding 
discussion. In this case, instead the predefined induced values of the fields on corresponding boundaries, some dynamical transition from $\mathscr{ A}$ field to the 
$\mathscr{ B}$ must exist as result of a classical equations of motion for the gauge field. Namely, it is a problem of the existence of a classical two-valued boundary solution.
Such solutions are known in high-energy scattering, see \cite{TwoVal} for examples. This question we plan to investigate in an additional publication.

 The simplest from the possibilities we consider is a direct complexification of the manifold by complexification of the manifold's coordinates.  The additional phase, i.e. additional coordinate, is factorized in the equations in this case with the help of the parameter assumed to be extremely small at the present:
\beq\label{Con3}
a_{\phi}\,\propto\,\frac{l_{0}}{R_{0}}\,.
\eeq
As mentioned in the Introduction, the obvious choice of the lengths in the definition is $l_{0}$ as Planck length and $R_{0}$ a manifold's curvature.  A consequence of that is a factorization of the real and complex parts of the metric,
i.e. factorization of real and complex parts of the corresponding complex manifold. Namely,
the smallness of the parameter guarantees
that the complexification is important and not small only when both parameters are of the same order, i.e. when $a_{\phi}\,\propto\,1$. In this case we have to consider an eight dimensional manifold instead the four dimensional one. 
That situation is possible only at some extremal points of the manifold's evolution. Otherwise the complexification is pure small distance effect, i.e. effect of quantum gravity. 
Namely, the proposed mechanism allows to determine the small contributions of the complex phases of an eight dimensional manifold to the quantities of the present classical four dimensional world.
This is similar to the compatifications of the additional dimensions in the string theory but not quite the same of course. The main difference between these mechanisms of the account of the classically non-observable dimensions is the following. In the present framework the contribution of the additional dimensions is factorized and can be treated perturbatively if the almost flat manifold without strong gravity fields is considered, whereas in the string's approach the contribution of the additional dimensions is always present, we can not take it equal to zero. Therefore, the framework with small $a$ does not require the compatification of the additional phase dimensions, instead it provides a smallness of their contributions in any expressions which we can treat perturbatively with respect to the parameter when the parameter is small. 
 
 The proposed complexification of the manifold through the complex coordinates is interesting also from the point of view of the symmetry group of the manifold. Namely, having the Poincare group representations as determination of the rule of the  classification of the existing particles, the natural question is about the allowed transformation group of the new 4D complex manifold. For the
$a_{\phi}\,\propto\,1$ limit there are a plenty of possibilities for the group's generalization, see an example and discussions in \cite{Penrose,Plebanski}. Anyway the final step in the 
complex twistor construction is a projection of the extended group on the real slice endowed with Poincare  group symmetry. The proposed case with small $a_{\phi}$ we treat differently.
First of all, the $a_{\phi}\,\propto\,0$ limit is well defined and determines a restoration of the Poincare symmetry. Secondly, considering the complex coordinates and expanding them with respect to 
$a_{\phi}$
we will obtain small corrections to the proposed classical transformations. Formally it means that to this precision it is enough to replace the real coordinates on the complex ones in the expressions for the group's representations and algebra of the real manifold and expand them in respect to $a_{\phi}$. In this case some quantum corrections will arise in the expressions of interests. 
Still, the symmetry group for the initial manifold can be any corresponding to the complex Minkowski space symmetry, see \cite{Penrose,Plebanski}.
This symmetry in the framework will restore at the $a_{\phi}\,\propto\,1$ limit, of course in this case the perturbative expansion can not be used. Therefore, in the proposed framework we discuss
not the projection of the complex manifold on the real slice, but a small complex corrections to the real metric, i.e. we consider a general framework with Lorentzian and different signatures metrics coexisting\footnote{The twistor space, definitely, as well describes manifolds endowed with metrics with different signatures, but it is not clear if it can be formulated as a dynamical model with simultaneous inclusion of metrics of different signatures.}. In this approach the metric with Lorentz signature is defined as a classical one and the metrics with other signatures contribute only at some special condition or at the quantum level.

 In any case, the complexification of the gravity action by the complexification of the coordinates results in the additional part to the "bare", real Einstein-Cartan action. This additional part provides 
a complex part to the classical "bare"' vierbein and consequently a complex additional part to the usual metric. For this type of the complexification we need to separate two cases. When we introduce a global
phase factor for the coordinates then the additional metric's part is a complex fluctuation above the usual metric, see  \eq{GSym13}. In general, for the non-expanded with respect to $a_{\phi}$ metric,
the action is a functional of the complex \eq{GSym2} Lagrangian, that in fact is not unusual, see  \cite{Witten} and \cite{Turok}. The interesting question, therefore, is a proper definition and properties of such complex action in the path integral, see discussions in \cite{Turok}. Introducing a local complex phase in the defintion of the complex coordinates, see  \eq{LSym1}-\eq{LSym2}, we introduce new gauge fields and their symmetry group G, in this case the $a_{\phi}$ parameter can be considered as a coupling constant of the group. This complexification of the manifold is more complicated than the first one of course, there is a possibility to obtain again complex fluctuations above the real metric, see \eq{VC20},  but, additionally, there is a possibility to introduce a metric with non-determinded signature from the very beginning with the help of the \eq{LSym1} $M$ field. This last case is  non-perturbative and complicated. We do not discuss it much in the paper postponing it for an additional publication. 

 There are following interesting properties of the actions \eq{GSym2} and \eq{LSym3} we obtained. First of all, there is no preferable axis of time direction, the metric's component can be of any sign in the situation with an undefined signature and any coordinate can serve as the time coordinate therefore. Moreover, fixing the metric, as usual Lorentzian for example, we still have a freedom to change the phases of the metric's components, i.e. to rotate the coordinate axes determining infinitely many ways of a foliation of the space-time. Each of these possibilities is described by the same action and, therefore, provides the same physics. In this case the preferable signature can be given by a  random selection from the infinitely many possibilities or by some fixation procedure similar to some extent to the spontaneous symmetry breaking. The later is possible when we talk about the  complexification by the gauge fields. Namely, in this case there is a possibility to define the classical values of the fields as a projection of some predefined boundary  fields. The approach is described in Appendix \ref{AppB} and the procedure provides a signature of the bulk by the value of the gauge fields on the boundaries of the manifold. The mutual property of all the actions with gauge fields involved is an appearance of the new factor in the front of the actions which is
a determinant of the gauge fields. The value of the factor 
is determined by the  boundary values of the fields and defines the relative weight of the action in the corresponding generating functional, otherwise it is arbitrary. The dynamics of such complex systems with many different parts of the general action in the generating functional is not clear and requires an additional investigation.

 A different way to introduce the dynamical signature of the metric is a generalization of the tangent space and an introduction of an
additional, auxiliary, metric in the tangent space which makes the tangent space curved. This can be achieved by the complexification of the vierbein with the help of the gauge fields, see \eq{VC02} and
\eq{VC021}, or by the direct definition of the usual vierbein fields as projection of some "gauge" vierbein performed by the gauge field of some symmetry group G, see \eq{V01}.
In both cases the final signature is dynamical and determined directly by the gauge fields, see \eq{VC22}, or by scalar fields and gauge fields together, see \eq{Tn12}. Again, for these mechanisms the projection procedure of Appendix \ref{AppB} is important. Without it the dynamics and correspondingly the signature can be arbitrary. Namely, as in the previous cases, there is neither preferable geometrical time nor preferable spatial coordinates and the given and only foliation in the approaches must be fixed separately, if required. We did not consider a matter issue in the frameworks,
see \cite{Kon} for the discussion about the possibilities of a proper definition of matter fields for the manifolds with complex metric.
It is interesting to understand in general how the  quantum matter fields behave in respect to the change of signature of the manifold and if there exists some dynamical mechanisms which relate
a foliation of the space-time and it's signature with properties of the matter. This problem requires an additional research and clarification of the properties and definition of the matter fields in respect to the manifold's symmetries and signature.

 Some interesting questions we can ask are about an existence of the different time's arrows directions in the manifolds with different signatures and corresponding issues related to it. First of all,
we note that if we stay in the framework of a perturbative approach with respect to the $a_{\phi}$ parameter, the possible additional contributions to any quantity of interests 
are extremely small. There is only one time's arrow on the classical level. 
Namely, the additional contributions are effectively pushed in the region of the quantum gravity regime, therefore any statements about the behavior of the time's arrow at this scale  must operate with a quantum gravity theory which we have no. Nevertheless, if we assume that the proposed approach correctly describes the quantum gravity regime or at least some of it's details, we can conclude that on this quantum level it possible that there is no any preferable time or spatial directions, unless some mechanisms fix the corrections to the metric as real with preferable signature. 
Considering the causality as a definition of the form of the corresponding propagators, we have no any problems with that till we do not consider some special regimes when $a_{\phi}$ is not small. In turn, an arbitrary value of the $a_{\phi}$ means the non-perturbative calculations  for the eight dimensional manifold with an arbitrary metric's tensor which has complex components.
The definition of the propagators in this case and their reduction to the usual ones
is very interesting question which we hope to explore in the future. 

 Another important problem is about a co-existing of the different regions with
different signatures and possible different time's arrows directions outside the perturbative regime. In this case we have two possibilities. 
The first one was considered in \cite{BKink}, there a case without the complex coordinates and/or metric
was discussed with some separation arises between the Euclidean and Lorentzian regions in a form of a hypersurface. 
In this set-up the hypersurface plays a role of a domain wall which separates the regions with different signatures and, in some extend, 
it defines an initial or final singularities for the time's arrows, see details in \cite{BKink}. As it seems, such hypersurfaces are unavoidable in the situation with co-existing of real metrics with different signatures, see also \cite{BonZub,IndG}. From the point of view of QFT, it can be considered also as 
Lorentzian space-time $\leftrightarrow$ Euclidean space-time geometrical transition vertices between separated parts of some mutual manifold. Due the discontinuities of the 
Einstein's tensor components on the hypersurface of separation, we again have no any problems with causality. We have two classically disconnected regions of space-time which possible connection
can be perhaps established only on the level of quantum gravity effects.
More complicated picture arises when we allow the complexification of the metric through some mechanisms. In this case we have no separating hypersurface between regions with different signatures due the complex phases
of the metric. In this situation we again obtain some eight dimensional manifold with two, or more, different time coordinates exist simultaneously, if the time's arrows can be 
defined for the metric's tensor with arbitrary complex components of course.
The notion of the causality in this case and possible mechanisms of the reduction of the manifold to the usual four dimensional one with one time's arrow are complicated problems which requires an additional investigation.

 Discussing the applications of the proposed approaches we note that they can be useful in an investigation of different aspects of the topology transition in both quantum gravity and cosmology through
\eq{Con2} expression for example.
In the paper we discussed a few possible mechanism of the \eq{Con1} effective action construction. It is interesting to understand which one can be realized in the nature. For that we need to understand the dynamics of the models with matter fields included.
Namely,
there is an interesting problem to determine the form of the action for the spinor of scalar fields in the new curved spaces of different signatures and investigate the dynamics of these
fields in corresponding models, see different aspects of this problem in  \cite{Dray,Viss,Barv,borde,Kri,SigC,Kon} references. Another interesting application of the dynamical signature is a clarification of it's possible correspondence to the new approaches to the classical gravity introduced and discussed at the last decade, see for example 
\cite{Bond,Villata,Chardin,Petit,Hoss,Kofinas,CPT,Sym,Linde,Kamen,Dayan}. We find an investigations of these ideas and possibilities very 
interesting.

\appendix
\newpage
\section{ Complex metric through complex vierbein}\label{AppA}
\renewcommand{\theequation}{A.\arabic{equation}}
\setcounter{equation}{0}

In order to illustrate how the \eq{V01} and \eq{V04} construction reproduce the \cite{Green} set-up, we firstly can consider as example the action of the following two-dimensional fixed unitary matrix
\beq\label{V08}
M = 
\begin{pmatrix}
1 & 0 \\
0 & \imath 
\end{pmatrix}
\,,\,\,\,
\tilde{M}\,=\,M^{T\,*} = \left( \begin{array}{c c}
1 & 0 \\
0 & -\imath 
\end{array} \right )\,,\,\,\,
\tilde{M}\,M\,=\,1\, \\
\eeq
on flat metric:
\begin{displaymath}\,
M\,
\left( \begin{array}{c c}
1 & 0 \\
0 & -1 
\end{array} \right )\,
M\,=\,
\left( \begin{array}{c c}
1 & 0 \\
0 & 1 
\end{array} \right )\,.
\end{displaymath}
Therefore, considering as $G$ the $U(4)$ group for example, we will obtain complex phases for the metric's components. Namely,
consider the spectral decomposition for the unitary matrix
\beq\label{V11}
\,M^{\,a \alpha}\,= \,\sum_{i=1}^{4}\,\lambda_{i}\,u^{a}_{i}\,\tilde{u}^{\alpha}_{i} 
\eeq
with $\lambda$ and $u$ as eigenvalues and eigenvectors, we obtain for the metric:
\beq\label{V12}
g_{\mu \nu}\,=\,\frac{1}{2}\,\sum_{i,j=1}^{4}\,\lambda_{i}\,\lambda_{j}\,\Le u^{\,a}_{i} \,\eta_{S\,a b}\,u^{b}_{j}\,\Ra\,\tilde{u}^{\,\alpha}_{i}\,
\tilde{u}^{\beta}_{j}\, \Le e_{\,\mu \alpha}\,e_{\,\nu \beta}\,+\,e_{\,\nu \alpha}\,e_{\,\mu \beta}  \Ra\,.
\eeq
Defining the local set of the vierbein through the identities
\beq\label{V13}
e_{\,\mu \alpha}\,\tilde{u}^{\,\alpha}\,_{i}\,=\,\delta_{\,\mu\,i}
\eeq
we will have finally
\beq\label{V121}
g_{\mu \nu}\,=\,\frac{1}{2}\,\lambda_{\mu}\,\lambda_{\nu}\,\Le\,
u^{\,a}_{\mu} \,\eta_{S\,a b}\,u^{b}_{\nu}\,+\,u^{\,a}_{\nu} \,\eta_{S\,a b}\,u^{b}_{\mu}\,\Ra\,.
\eeq
Now, if  we restrict ourselves by the diagonal unitary matrices, the corresponding eigenvectors are real and
orthonormal. Therefore, for the arbitrary four dimensional diagonal unitary matrix
\beq\label{V131}
M^{a \alpha} = 
\begin{pmatrix}
e^{\imath\,\alpha_1} & 0 & 0 & 0 \\
0 & e^{\imath\,\alpha_2} & 0 & 0 \\
0 & 0 & e^{\imath\,\alpha_3} & 0 \\
0 & 0 & 0 & e^{\imath\,\alpha_4}
\end{pmatrix}
\eeq
we obtain a simple expression for the generalized flat metric:
\beq\label{V14}
g_{\mu \nu}\,=\,e^{\,\imath\,(\alpha_{\mu}\,+\alpha_{\nu})}\,\eta_{S\,\mu\,\nu}\,.
\eeq
We see, that in terms of \eq{V01} transform we simply can write the vierbein transform as
\beq\label{V15}
\,e^{\,a}_{\,\mu}\,=\,e^{\,\imath\,\ph_{a}}\,\delta^{\, a \alpha}\,e_{\,\mu\, \alpha}\,
\eeq
obtaining for the metric
\beq\label{V16}
g_{\,\mu \nu}\,=\,\,e^{\,\imath (\phi_{a}\,+\,\phi_{b})}\eta_{S\,a b}\,e^{\,a}_{\,\mu}\,e^{\,b}_{\,\nu}\,
\eeq
which describes, at a first sight, a metric with indefinite complex signature. Nevertheless, we remind that the $\phi$ angles are dynamical fields in the approach,
therefore the final leading order expression for the metric will be determined by the classical values of these fields.

\newpage
\section{ Induced part of the action}\label{AppB}
\renewcommand{\theequation}{B.\arabic{equation}}
\setcounter{equation}{0}

 Following \cite{EffA,EffA1} we consider the action given in \eq{VC10}. For this action in the flat-space time we have:
\beq\label{VC16}
S_{A}\,=\,-\,\frac{1}{4}\int\,d^4 x\,
tr \left[ G_{\mu \nu}\,G^{\mu \nu} \right]\,
\eeq
and for the induced part of the action 
\beq\label{VC17}
S_{ind}\,=\,-\,\sum_{i}\,\int\,d^4 x\,tr\left[ \,\Le \D_{\mu} O(A_{\mu})\Ra \,\Le \D_{\nu}^{2} \mathscr{ A}^{\mu}_{i} \Ra\,\right]\,,
\eeq  
where the Riemann indexes are summed up through the Minkowski metric as usual. The $\mathscr{ A^{\mu}}$ fields are
defined at the boundaries, and they intend to provide the signature of the additional part in the \eq{VC03} metric. For example,  we can define the two complete sets of the  boundary fields 
which satisfy
\beq\label{VC20}
\D_{\mu}\,\mathscr{ A}^{\mu}_{i}\,=\,0\,
\eeq
and the following boundary conditions:
\beq\label{VC18}
\left\{
\begin{array}{clr}
\mathscr{ A}^{\,\mu}_{1}(x)\,\rightarrow\,0 & x^{0}\,\rightarrow\,\infty\,, \\
\mathscr{ A}^{\,\mu}_{2}(x)\,\rightarrow\,0 & x^{0}\,\rightarrow\,-\,\infty\,.
\end{array} \right.
\eeq
Also the following term must be added to the action
\beq\label{VC2201}
S_{\mathscr{ A}}\,=\,\sum_{i}\,\int\,\mathscr{ A}^{\,\mu}_{i}\,\D_{\nu}^{2}\,\mathscr{ A}_{\,\mu\,i}\,
\eeq
which preserves the correct form of the propagators in the full action, see discussions in \cite{EffA,EffA1}.
Therefore, for the gauge fields 
\beq\label{VC19}
\D_{\mu}\, A^{\mu}\,=\,0\,
\eeq
we obtain as a solution of the equations of motion\footnote{Following the analogy with the high energy scattering approach, we can consider the \eq{VC16} action with an additional induced term 
as describing a "scattering" between two boundary fields with boundaries defined at the edges of time.}:
\beq\label{VC21}
A^{\mu}_{\,cl}\,=\,\mathscr{ A}^{\,\mu}_{1}\,+\,\mathscr{ A}^{\,\mu}_{2}\,.
\eeq
The operator $O$ in the \eq{VC17} action is defined similarly to definitions of \cite{EffA,EffA1}. In the simplest variant it is 
\beq\label{VC23}
O(A_{\mu})\,=\,\frac{1}{a_{\phi}\,C(R)}\,P\, e^{a_{\phi}\,\int_{-\infty}^{x^{\mu}}\,dx^{'\mu}\, A_{\mu}(x^{'})} \,.
\eeq
There is no summation on $\mu$ index in the ordered exponential and the index is fixed in correspondence to the \eq{VC17} expression, 
$C(R)$ is the eigenvalue of Casimir operator in the representation R for the chosen gauge symmetry group.
The different form of this operator and discussion about can be found in \cite{EffA,EffA1}.

\end{document}